%Paper: hep-th/9401008
%From: ghoshal@physics.rutgers.edu (Subir Ghoshal)
%Date: Tue, 4 Jan 94 21:54:47 EST

\input harvmac
\rightline{RU-94-02}
\vskip 1 in
\centerline{\bf Boundary S-matrix of the $O(N)$-symmetric Non-linear
Sigma Model}
\vskip .5 in
\centerline {Subir Ghoshal\foot{E-mail: GHOSHAL@ruhets.rutgers.edu}}
\centerline{Department of Physics and Astronomy}
\centerline{Rutgers University}
\centerline{P.O.Box 849, Piscataway, NJ 08855-0849}
\vskip 1 cm
\centerline{\bf Abstract}
We conjecture that the $O(N)$-symmetric non-linear sigma model in the
semi-infinite $(1+1)$-dimensional space is ``integrable'' with respect
to the ``free'' and the ``fixed'' boundary conditions. We then
derive, for both cases, the boundary S-matrix for the reflection of massive
particles
of this model off the boundary at $x=0$.

\lref\zaza{A.B.Zamolodchikov,Al.B.Zamolodchikov.Ann.Phys.120(1979),253.}
\lref\ghza{S.Ghoshal and A.B.Zamolodchikov,''Boundary S-matrix and
Boundary State in Two-dimensional Integrable Quantum Field
Theory'',Rutgers Preprint,RU-93-20,hep-th/9306002.}
\lref\poly{A.M.Polyakov, Phys.Lett.B 59 (1975) 87.}
\lref\thun{M.Karowski, H.J.Thun, T.T.Truong and P.H.Weisz, Phys.Lett.B 67
(1977), 32.}
\lref\dveg{H.J.De Vega,A.Gonzalez Ruiz, Journal of Physics, A26, 12, L519.}
\lref\frko{A.Fring, R.Koberle. ``Factorized Scattering in the Presence
of Reflecting Boundaries''. Preprint USP-IFQSC/TH/93-06, 1993.}
\lref\chrk{I.Cherednik. Theor.Math.Phys.,61,35 (1984) p.977.}
\lref\skly{E.K.Sklyanin. Funk.Analiz Prilozh. 21 (1987) p86.}

\newsec{\bf Introduction}
 Recently, two-dimensional integrable field theory with a
 reflecting boundary has been studied \refs{\chrk,\skly,\dveg,\frko,\ghza}. An
essential ingredient of an integrable
 field theory in the infinite space is the existence of an infinite number
 of mutually commuting integrals of motion. In the presence of the boundary,
 for arbitrary boundary conditions, these ``charges'' no longer remain
 conserved. However, sometimes, it is possible to modify these charges
 with special ``integrable'' boundary conditions so that the modified
 charges are indeed conserved. Then such a theory may be called a
 ``two-dimensional integrable field theory with a boundary''.
 \par An integrable ``bulk'' field theory enjoys the property that its
    multi-particle S-matrix amplitude factorizes into a product of an
    appropriate number of two-particle S-matrix amplitudes. The latter
    satisfy several constraints, namely, Yang-Baxter equation,
    unitarity and crossing symmetry \zaza. These constraints enable one to
    compute the exact S-matrix up to the so-called ``CDD''-factors. It has
    been known for quite some time how to generalise this factorizable
    structure of the S-matrix in the presence of a reflecting boundary
    \chrk. In addition to the ``bulk'' two-particle S-matrices one needs
    to introduce specific ``boundary reflection amplitudes'' describing
    reflections of various particles in the theory when they fall on the
boundary. The latter have to
    satisfy appropriate generalisations of the constraints of the bulk
    theory $-$ the ``boundary Yang-Baxter equation'', the ``boundary unitarity
    condition'' and the ``boundary cross-unitarity condition''. The last of
    these was introduced in \ghza. Thus, one can in a way similar to the
    bulk case, pin down the factorizable boundary S-matrix, again, up to the
``CDD''-factors.
\par In this paper we study the boundary S-matrix of the
$O(N)$-symmetric non-linear sigma model. In section-$2$, we briefly
introduce the model and conjecture that the model is integrable in the
semi-infinite space with the ``free'' and the ``fixed'' boundary conditions. In
section-$3$, we study the fully $O(N)$-symmetric
boundary scattering which is the case for the ``free'' boundary
conditions. In section-$4$, we consider the boundary scattering with the
``fixed''
boundary condition which enjoyes $O(N-1)$-symmetry. Concluding remarks
appear in section-$5$.

\newsec{\bf $O(N)$-symmetric Nonlinear Sigma Model}
 Let us collect some known facts about the $O(N)$-symmetric nonlinear
 sigma model in infinite $(1+1)$-dimensional space. It is described by the
Lagrangian density and the
 constraint \eqn\lagrangian{{\cal L}={1\over  2
 g_{0}}\sum_{i=1}^{N}(\partial_{\mu}n_{i})^{2}; \quad \quad \quad
 \sum_{i=1}^{N}n_{i}^2=1 } where $n_i=n_{i}(x,y)$ are $N$ scalar fields with
 $N=3,4,...$ and $g_0$ is a (bare) coupling constant. The model is
 $O(N)$-symmetric, renormalizable and asymptotically free \refs{\poly,\thun}.
It has $N$
 massive particles in the $O(N)$-multiplet. The exact S-matrix
 of this model has been found in \zaza. The scattering involves no
 particle production and the $n$-particle S-matrix factorizes into a product of
two particle S-matrices. So, it is
 sufficient to obtain the two-particle scattering amplitude. Let us
 parametrize the energy and momentum of a particle of mass $m$ in terms
 of the rapidity variable $\theta$ in the usual manner
 \eqn\rapidity{p^{0}=m\, cosh\theta \kern .5 cm ; \kern .5 cm p^{1}=m\,
 sinh\theta }Then the
 following commutation relation between the formal particle creating
 operators $A_{i}(\theta)$ describes the two-particle scattering:
\eqn\twopartsm{A_{i}(\theta)A_{j}(\theta')=\delta_{ij}\sigma_{1}(\theta-\theta')\sum_{k=1}^{N}A_{k}(\theta')A_{k}(\theta)+\sigma_{2}(\theta-\theta')
A_{j}(\theta')A_{i}(\theta)+\sigma_{3}(\theta-\theta')
A_{i}(\theta')A_{j}(\theta)} where
 \eqn\bulksmone{\sigma_{1}(\theta)=-{i\lambda \over i\pi -
 \theta}\sigma_{2}(\theta)}
 \eqn\bulksmthree{\sigma_{3}(\theta)=
 -{i\lambda \over \theta }\sigma_{2}(\theta) }
 \eqn\bulksmtwo{\sigma_{2}(\theta)={\Gamma ({\lambda \over
 2\pi}-i{\theta \over 2\pi})\Gamma ({1\over 2}-i{\theta \over
 2\pi})\over \Gamma({1\over 2}+{\lambda \over 2\pi}-i{\theta \over 2
 \pi})\Gamma (-i{\theta \over 2\pi})}    {\Gamma ({1\over 2}+{\lambda \over
 2\pi}+i{\theta \over 2\pi})\Gamma (1+i{\theta \over
 2\pi})\over \Gamma(1+{\lambda \over 2\pi}+i{\theta \over 2
 \pi})\Gamma ({1\over 2}+i{\theta \over 2\pi})}} and
 \eqn\generallamb{\lambda = {2\pi \over N-2}}The $O(N)$-sigma model in
 the semi-infinite space with a boundary at $x=0$ can be described by the
 following action \eqn\bdaction{ \int_{-\infty}^{\infty}dy
 \int_{-\infty}^{0}dx\; {\cal L}\; +\;\int_{-\infty}^{\infty} dy \; {\cal
L}_{B}} where $ {\cal L}_{B}$, the
boundary Lagrangian density, encodes the boundary conditions for the theory. In
general ${\cal L}_{B}$ may spoil the integrability of the original field
theory because under general boundary conditions one does not expect the
integrals of motion of the original field theory to survive in the
boundary field theory. However, under specific choices of ${\cal L}_{B}$
these integrals of motion, although modified, continue to remain so. In
that case we have a ``boundary integrable field theory''$\,$\ghza.  In
this paper we consider two simple boundary conditions
\eqn\freebd{\hbox{1. Free
Boundary Condition:\quad }  {\cal
L}_{B}=0 \kern 5 cm} \eqn\fixedbd{\hbox{2. Fixed Boundary Condition:\quad}
n_{i}(x,y)|_{x=0}=n_{i}^{(0)};\quad
n_{i}^{(0)}=(0\, 0\, 0\,....1) } The fixed boundary condition can also be
recovered as the $h\rightarrow\infty$ limit of \eqn\bdfield{{\cal
A}_{B}=\int_{-\infty}^\infty dy \;h_{i}n_{i}\quad\quad ;\quad \quad h_{i}=h(0\,
0\, 0\,... 1)} One
has to examine the integrals of motion of the $O(N)$-sigma model in order to
claim integrability of the boundary field theory with the ``free'' or
the ``fixed'' boundary conditions. Here, we simply conjecture that both of
these boundary conditions preserve integrability.
\par In addition to the bulk S-matrices we need to consider the boundary
reflection amplitudes in the boundary theory. The amplitudes for
reflections of $N$ species of particles off the boundary can be derived
using the boundary
bootstrap approach as discussed in \refs{\ghza,\frko}. The reflection of a
particle
of the $i$-th kind (Fig.1) is formally encoded in
\eqn\reflection{A_{i}(\theta)B=R_{i}^{j}(\theta)A_{j}(-\theta)B} where
$A_{i}(\theta)$ and $B$ formally represent the operators that create the
$i$-th particle and the boundary respectively.
 Using \reflection $\,$
repeatedly one can obtain the ``boundary unitarity condition''(Fig.2)
\eqn\unitarity{R_{i}^{j}(\theta)R_{j}^{k}(-\theta)=\delta_{i}^{k}}where
$\delta_{i}^{j}$ is the well-known Kronekar delta simbol. The
``boundary cross-unitarity'' equation was introduced in \ghza. It reads
\eqn\gencross{R_{i}^{j}(i{\pi \over
2}-\theta)=S^{ij}_{kl} (2\theta) \, R_{k}^{l}(i{\pi \over 2}+\theta)}where
$S_{ij}^{kl} (2\theta)$ is the two particle scattering amplitude for
$i+j\rightarrow k+l$ scattering. In addition, the boundary scattering
amplitudes must also satisfy the ``boundary Yang-Baxter equation''
(illustrated in Fig.3)
\eqn\bybeq{R_{j}^{m}(\theta')S_{im}^{np}(\theta+\theta')R_{n}^{q}(\theta)S_{pq}^{lk}(\theta-\theta')=
S_{ij}^{mn}(\theta-\theta')R_{m}^{p}(\theta)S_{np}^{qk}(\theta+\theta')R_{q}^{l}(\theta')}

\newsec{\bf Boundary S-matrix For The Free Boundary Condition}
In this case the boundary S-matrix must enjoy full
$O(N)$-symmetry and is therefore described by the following ansatz:
\eqn\ONsymm{R_{i}^{j}(\theta)=\delta_{i}^{j}R(\theta)}Then the
boundary unitarity condition \unitarity $\,$and the boundary
cross-unitarity condition \gencross $\,$for this case assume the
following forms respectively \eqn\ONunitarity{R(\theta)R(-\theta)=1}
\eqn\bdcrossing{R(i{\pi \over
2}-\theta)=\sigma (2\theta) R(i{\pi \over 2}+\theta)}

\vfill
\eject

where

\eqn\generalsig{\sigma
(\theta)=N\sigma_{1}(\theta)+\sigma_{2}(\theta)+\sigma_{3}(\theta)}or
using \bulksmone,\bulksmthree, and \bulksmtwo

\eqn\fullsigma{\sigma(\theta)=-{\Gamma (1+{\lambda \over
 2\pi}-i{\theta \over 2\pi})\Gamma (-{1\over 2}-i{\theta \over
 2\pi})\over \Gamma(1+{\lambda \over 2\pi}+i{\theta \over 2
 \pi})\Gamma (-{1\over 2}+i{\theta \over 2\pi})}    {\Gamma ({1\over
2}+{\lambda \over
 2\pi}+i{\theta \over 2\pi})\Gamma (1+i{\theta \over
 2\pi})\over \Gamma({1\over 2}+{\lambda \over 2\pi}-i{\theta \over 2
 \pi})\Gamma (1-i{\theta \over 2\pi})}}

The boundary Yang-Baxter equation \bybeq $\,$ turns out to be an identity for
this fully $O(N)$-symmetric case. So,
\ONunitarity $\,$and \bdcrossing $\,$constitute the set of equations that
enable us
to pin down the boundary reflection amplitudes. The solution is
\eqn\fullbdsm{R(\theta)={\Gamma ({1\over 2}+{\lambda \over
 4\pi}-i{\theta \over 2\pi})\Gamma (1+i{\theta \over
 2\pi})\over \Gamma({1\over 2}+{\lambda \over 4\pi}+i{\theta \over 2
 \pi})\Gamma (1-i{\theta \over 2\pi})}    {\Gamma ({3\over 4}+{\lambda
 \over 4\pi}+i{\theta
 \over 2\pi})\Gamma ({1\over 4}-i{\theta \over
 2\pi})\over \Gamma({3\over 4}+{\lambda \over 4\pi}-i{\theta \over 2
 \pi})\Gamma ({1\over 4}+i{\theta \over 2\pi})}}There are no poles in
$R(\theta)$ in the
``physical strip'' $0 < \theta < i{\pi\over 2}$ . This is consistent
with the fact that the repulsive
 interaction in the bulk theory does not allow any bound states. Any
 such bound state of the bulk theory would contribute a pole in
 $R(\theta)$$\,$\ghza. Additional poles might still occur in $R(\theta)$ due to
boundary bound
 states. Because we do not have any particular reason to expect such
 boundary bound states, we conjecture that \fullbdsm, without any
 additional ``CDD'' factors describes the boundary scattering for the
 ``free'' boundary condition.

\newsec{\bf Boundary S-matrix For The Fixed Boundary Condition}
For the fixed boundary condition we do not expect the boundary
scattering to respect the full $O(N)$-symmetry. However, with the field
on the boundary pointing in a certain direction we have a
$O(N-1)$-symmetric boundary scattering. Consistent with this symmetry the
ansatz for the boundary
S-matrix is
$$R_{i}^{j}(\theta)=\delta_{i}^{j}R_{1}(\theta);$$
$$R_{N}^{N}(\theta)=R_{2}(\theta)$$
\eqn\NOsymm{R_{i}^{N}(\theta)=R_{N}^{i}(\theta)=0; \quad \quad \quad
i,j=1,2,3,...(N-1)}

\vfill
\eject

The  Boundary Yang-Baxter equation \bybeq $\,$for this case results in two
independent equations:
$$R_{2}(\theta')R_{1}(\theta)\sigma_{1}(\theta+\theta')[(N-1)\sigma_{1}(\theta-\theta')+\sigma_{2}(\theta-\theta')+\sigma_{3}(\theta-\theta')]$$$$+R_{2}(\theta')R_{2}(\theta)\sigma_{1}(\theta-\theta')[\sigma_{1}(\theta+\theta')+\sigma_{2}(\theta+\theta')+\sigma_{3}(\theta+\theta')]$$
$$=R_{1}(\theta')R_{1}(\theta)\sigma_{1}(\theta-\theta')[(N-1)\sigma_{1}(\theta+\theta')+\sigma_{2}(\theta+\theta')+\sigma_{3}(\theta+\theta')]$$
\eqn\bYBone{
+R_{1}(\theta')R_{2}(\theta)\sigma_{1}(\theta+\theta')[\sigma_{1}(\theta-\theta')+\sigma_{2}(\theta-\theta')+\sigma_{3}(\theta-\theta')]}

and
$$R_{2}(\theta')R_{1}(\theta)\sigma_{2}(\theta+\theta')\sigma_{3}(\theta-\theta')+R_{2}(\theta')R_{2}(\theta)\sigma_{2}(\theta-\theta')\sigma_{3}(\theta+\theta')$$
\eqn\bYBtwo{
=R_{1}(\theta')R_{2}(\theta)\sigma_{2}(\theta+\theta')\sigma_{3}(\theta-\theta')+R_{1}(\theta')R_{1}(\theta)\sigma_{2}(\theta-\theta')\sigma_{3}(\theta+\theta')}

Solving these equations we get
\eqn\amplratio{{R_{1}(\theta)\over R_{2}(\theta)}={i\pi \pm 2\theta\over
i\pi + 2\theta}}

The $(+)$ sign in the numerator implies $R_{1}(\theta)=R_{2}(\theta)$
which is the case of full $O(N)$-symmetry discussed in the previous
section. The $(-)$ sign distinguishes between $R_{1}(\theta)$ and
$R_{2}(\theta)$
and therefore gives the solution for the present case. So, we can now write
\eqn\amplratiof{{R_{1}(\theta)\over R_{2}(\theta)}={i\pi-2\theta\over
i\pi +  2\theta}}

The boundary unitarity equation \unitarity $\;$ in the present case
translates into \eqn\partunitarity{R_{1}(\theta)R_{1}(-\theta)=1 \kern .5
cm ;\kern .5
cm R_{2}(\theta)R_{2}(-\theta)=1}   while the boundary cross-unitarity equation
\gencross
$\,$becomes \eqn\crossingRone{R_{1}(i{\pi\over 2}-\theta)=R_{1}(i{\pi\over
2}+\theta)[(N-1)\sigma_{1}(2\theta)+\sigma_{2}(2\theta)+
\sigma_{3}(2\theta)]+R_{2}(i{\pi\over
2}+\theta)\sigma_{1}(2\theta)}\eqn\crossingRtwo{R_{2}(i{\pi\over
2}-\theta)=R_{2}(i{\pi\over 2}+\theta)[\sigma_{1}(2\theta)+\sigma_{2}(2\theta)+
\sigma_{3}(2\theta)]+R_{1}(i{\pi\over 2}+\theta)(N-1)\sigma_{1}(2\theta)}
Using \amplratiof,\bulksmone,\bulksmtwo $\,$and \bulksmthree $\,$ we see that
\crossingRone $\,$
$\,$and$\,$\crossingRtwo $\,$ both reduce to the same following equation,
indicating the consistency of the boundary bootstrap
approach:\eqn\crossingR{R_{1}(i{\pi\over
2}-\theta)= -{i{\lambda\over
2}-\theta\over i{\lambda\over 2}+\theta}\sigma(2\theta)R_{1}(i{\pi\over
2}+\theta)}where $\sigma(\theta)$ is the same as in \fullsigma. We now have to
solve \partunitarity$\;$  and \crossingR$\;$
simultaneously to obtain $R_{1}(\theta)$.
The solution reads \eqn\solRone{ R_{1}(\theta)={\Gamma ({1\over 2}+{\lambda
\over
 4\pi}-i{\theta \over 2\pi})\Gamma (1+i{\theta \over
 2\pi})\over \Gamma({1\over 2}+{\lambda \over 4\pi}+i{\theta \over 2
 \pi})\Gamma (1-i{\theta \over 2\pi})}    {\Gamma ({1\over 4}+{\lambda
 \over 4\pi}+i{\theta
 \over 2\pi})\Gamma ({3\over 4}-i{\theta \over
 2\pi})\over \Gamma({1\over 4}+{\lambda \over 4\pi}-i{\theta \over 2
 \pi})\Gamma ({3\over 4}+i{\theta \over 2\pi})}}Using \amplratiof$\,$ we
 have for $R_{2}(\theta)$ then \eqn\solRtwo{ R_{2}(\theta)={\Gamma ({1\over
2}+{\lambda \over
 4\pi}-i{\theta \over 2\pi})\Gamma (1+i{\theta \over
 2\pi})\over \Gamma({1\over 2}+{\lambda \over 4\pi}+i{\theta \over 2
 \pi})\Gamma (1-i{\theta \over 2\pi})}    {\Gamma ({1\over 4}+{\lambda
 \over 4\pi}+i{\theta
 \over 2\pi})\Gamma (-{1\over 4}-i{\theta \over
 2\pi})\over \Gamma({1\over 4}+{\lambda \over 4\pi}-i{\theta \over 2
 \pi})\Gamma (-{1\over 4}+i{\theta \over 2\pi})}}There are no poles in
 $R_{1}(\theta)$ in the
``physical strip'' $0 < \theta < i{\pi\over 2}$ . For $R_{2}(\theta)$ the same
 is true except for the existence of a pole at $\theta=i{\pi\over 2}$.
 This pole signals the presence of the $1$-particle contribution in the
 corresponding boundary state \ghza. Clearly, the $O(N-1)$-symmetry of
 the ``fixed'' boundary condition allows the contribution of the
 state $|A_{N}>$ in this boundary state.

\newsec{\bf Conclusion}
In this paper we have considered the boundary scattering with the ``free''
and the ``fixed'' boundary conditions for the $O(N)$-symmetric non-linear
sigma model and conjectured the boundary S-matrix in each case. Of
course, our
conjecture that for both of these boundary conditions the boundary field
theory remains integrable will have to be supported by explicit
demostration that the integrals of motion for this theory do indeed
survive in the boundary theory.
\par Also it would be interesting to analyse, in general, what boundary
conditions for the $O(N)$-sigma model are consistent with integrability.
\vskip .5 in
\centerline{\bf Acknowledgement}
 I would like to thank A.B.Zamolodchikov for advice, inspiration and
 numerous illuminating discussions.

\listrefs
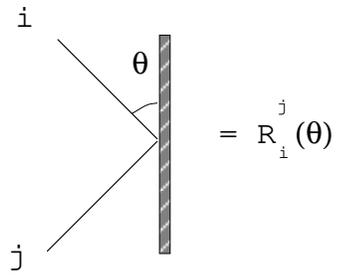
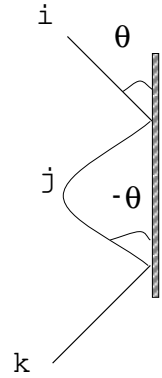
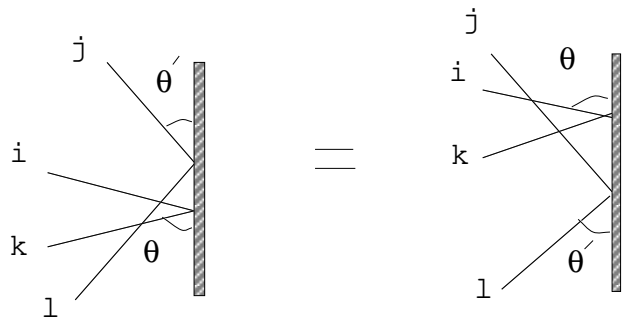
\end